\documentclass{KapProc} % Computer Modern font calls

\def\lsim{\mathrel{\rlap{\lower 4pt \hbox{\hskip 1pt $\sim$}}\raise 1pt \hbox
        {$<$}}}
\def\gsim{\mathrel{\rlap{\lower 4pt \hbox{\hskip 1pt $\sim$}}\raise 1pt \hbox
        {$>$}}}

\kluwerbib
\input{psfig}

\begin{document}

%------------ article title  ------------------->>

% If you use \\'s , please supply an alternate version of the title
% in square brackets, i.e., 
%\articletitle[Communism, Sparta, and Plato]
%{COMMUNISM, SPARTA,\\ and PLATO}

\articletitle{Hypernova Nucleosynthesis and  \\
Galactic Chemical Evolution}

\vspace{1cm}

\notes{\footnotesize To be published in "The Influence of Binaries on 
Stellar Population Studies", \\ 
\hspace*{0.35cm} ed. D. Vanbeveren (Kluwer), 2001.}

%% optional, to supply a shorter version of the title for the running head:
%%\chaptitlerunninghead{}

\author{Ken'ichi Nomoto, Keiichi Maeda, Hideyuki Umeda \\ 
 Takayoshi Nakamura}
\affil{Department of Astronomy and Research Center for the Early Universe\\
School of Science, University of Tokyo}

%\author{}
%\affil{Author Affiliation\\
%Second Line of Affiliation}

%% multiple authors may be separated with \\
%% \author{Samuel Bostaph\\
%% and Gregor Kariotis}

% optional prologue
%\prologue{To be published in "The Influence of Binaries on Stellar Population
%Studies", ed. D. Vanbeveren (Kluwer), 2001.}{}

% optional abstract
\begin{abstract}

We study nucleosynthesis in 'hypernovae', i.e., supernovae with very
large explosion energies ($ \gsim 10^{52} $ ergs) for both spherical
and aspherical explosions.  The hypernova yields compared to those of
ordinary core-collapse supernovae show the following characteristics:
1) Complete Si-burning takes place in more extended region, so that
the mass ratio between the complete and incomplete Si burning regions
is generally larger in hypernovae than normal supernovae.  As a
result, higher energy explosions tend to produce larger [(Zn, Co)/Fe],
small [(Mn, Cr)/Fe], and larger [Fe/O], which could explain the trend
observed in very metal-poor stars.  2) Si-burning takes place in lower
density regions, so that the effects of $\alpha$-rich freezeout is
enhanced.  Thus $^{44}$Ca, $^{48}$Ti, and $^{64}$Zn are produced more
abundantly than in normal supernovae.  The large [(Ti, Zn)/Fe] ratios
observed in very metal poor stars strongly suggest a significant
contribution of hypernovae.  3) Oxygen burning also takes place in
more extended regions for the larger explosion energy.  Then a larger
amount of Si, S, Ar, and Ca ("Si") are synthesized, which makes the
"Si"/O ratio larger.  The abundance pattern of the starburst galaxy
M82 may be attributed to hypernova explosions.  Asphericity in the
explosions strengthens the nucleosynthesis properties of hypernovae 
except for "Si"/O.  We thus suggest that hypernovae make important
contribution to the early Galactic (and cosmic) chemical evolution.

\end{abstract}

% optional keywords
% \begin{keywords}
% Text, text...
% \end{keywords}

%------------ body of article ------------------->>
\section{Introduction}

Massive stars in the range of 8 to $\sim$ 100$M_\odot$ undergo
core-collapse at the end of their evolution and become Type II and
Ib/c supernovae (SNe II and SNe Ib/c) unless the entire star collapses
into a black hole with no mass ejection (e.g., Arnett 1996).  These
SNe II and SNe Ib/c (as well as Type Ia supernovae) release large
explosion energies and eject explosive nucleosynthesis products, thus
having strong dynamical, thermal, and chemical influences on the
evolution of interstellar matter and galaxies.  Therefore, the
explosion energies of core-collapse supernovae are fundamentally
important quantities. An estimate of $E \sim 1\times 10^{51}$ ergs
has often been used in calculating nucleosynthesis (e.g., Woosley \&
Weaver 1995; Thielemann et al. 1996) and the impact on the
interstellar medium.  (In the present paper, we use the explosion
energy $E$ for the final kinetic energy of explosion.)

SN1998bw called into question the applicability of the above energy
estimate for all core-collapse supernovae.  This supernova was
discovered in the error box of the gamma-ray burst GRB980425, only 0.9
days after the date of the gamma-ray burst and was very possibly
linked to it (Galama et al. 1998).  SN1998bw is classified as a Type Ic
supernova (SN Ic) but it is quite an unusual supernova.  The very
broad spectral features and the light curve shape have led to the
conclusion that SN 1998bw was a hyper-energetic explosion of a massive
($\sim$ 14 $M_\odot$) C + O star with $E$ = 3 - 6 $ \times$ $10^{52}$
ergs, which is about thirty times larger than that of a normal
supernova (Iwamoto et al. 1998; Woosley et al. 1999;
Branch 2000; Nakamura et al.  2001a).  The amount of $^{56}$Ni
ejected from SN1998bw is found to be $M(^{56}$Ni) $\simeq$ 0.4 - 0.7
$M_\odot$ (Nakamura et al. 2001a; Sollerman et al. 2000), which is
about 10 times larger than the 0.07$M_\odot$ produced in SN1987A, a
typical value for core-collapse supernovae.

In the present paper, we use the term 'Hypernova' to describe such an
extremely energetic supernova as $E \gsim 10^{52}$ ergs without
specifying the nature of the central engine (Nomoto
et al. 2000).

Recently, other hypernova candidates have been recognized.  SN1997ef
and SN1998ey are also classified as SNe Ic, and show very broad
spectral features similar to SN1998bw (Iwamoto et al. 2000).
The spectra and the light curve of SN1997ef have been well simulated
by the explosion of a 10$M_\odot$ C+O star with $E = 1.0 \pm 0.2$
$\times$ $10^{52}$ ergs and $M(^{56}$Ni) $\sim$ 0.15 $M_\odot$
(Iwamoto et al. 2000; Mazzali, Iwamoto, \& Nomoto 2000).  SN1997cy is
classified as a SN IIn and unusually bright (Germany et al. 2000;
Turatto et al. 2000).  Its light curve has been simulated by a
circumstellar interaction model which requires $E \sim$ 5 $\times$
$10^{52}$ ergs (Turatto et al. 2000).  The spectral similarity of
SN1999E to SN1997cy (Cappellaro et al. 1999) would suggest that
SN1999E is also a hypernova.  Note that all of these estimates of $E$
assume a spherically symmetric event.

Hypernovae may be associated with the formation of a black hole as has
been discussed in the context of the GRB-SNe connection (Woosley 1993;
Paczynski 1998; Iwamoto et al. 1998; MacFadyen \& Woosley 1999).  In
these models, the gravitational energy, or the rotational energy would
be released via pair-neutrino annihilation or the Blandford-Znajek
mechanism (Blandford \& Znajek 1977).  Alternatively, large magnetic
energies are released from a possible magnetar (Nakamura 1998; Wheeler
et al. 2000).  The explosion may also be aspherical (H\"oflich et al.
1999; MacFadyen \& Woosley 1999; Khokhlov et al. 1999).

We investigate the characteristics of nucleosynthesis in such
energetic core-collapse hypernovae, the systematic study of which has
not yet been done.  We examine both spherical and aspherical explosion
models and discuss their contributions to the Galactic chemical
evolution.

\section {Spherical Hypernova Explosions}

We use various progenitor models with the main sequence masses of 20,
25, 30, and 40$M_{\odot}$ (whose He core masses are 6, 8, 10,
16$M_{\odot}$, respectively) (Nomoto \& Hashimoto 1988; Umeda, Nomoto,
\& Nakamura 2000), metallicity of $Z =$ 0 to solar, and explosion
energies of $E_{51} = E/10^{51}$ ergs $=$ 1 - 100 (Nakamura et
al. 2001b; Umeda \& Nomoto 2001).

\begin{figure}
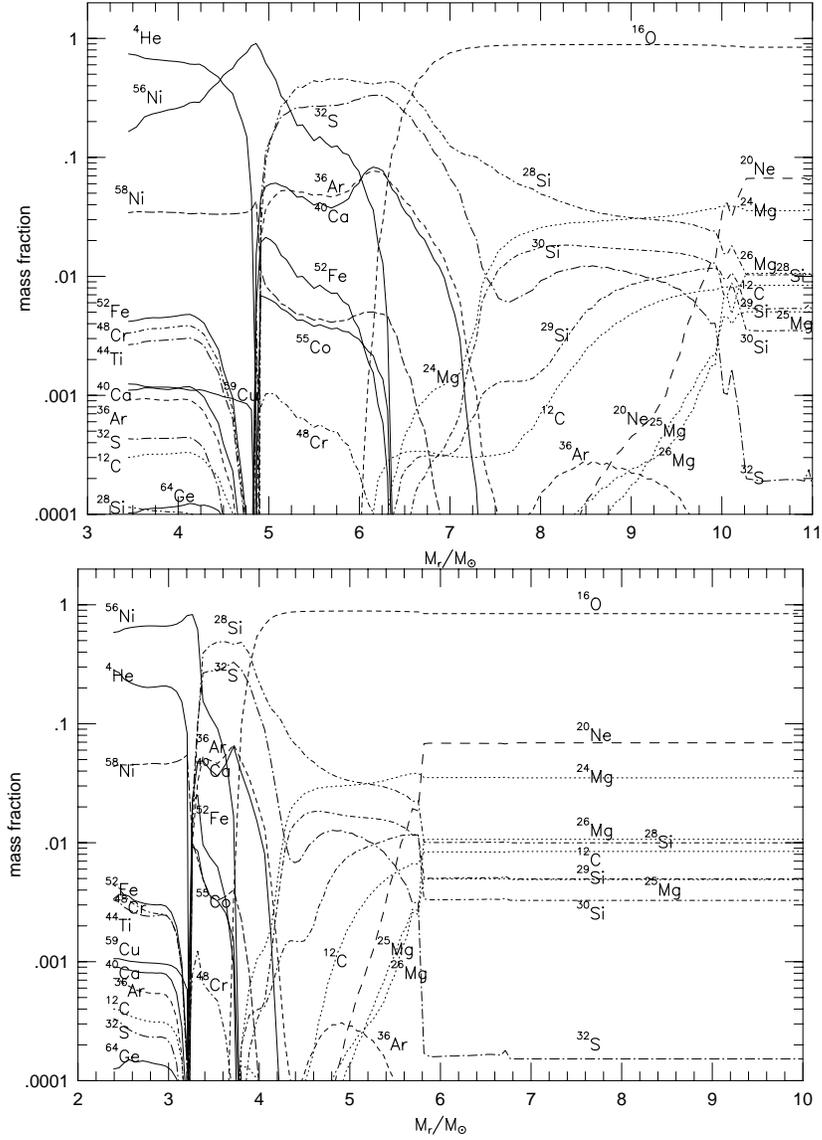

\begin{center}\leavevmode
\psfig{figure=fig2a.epsi,height=7.5cm}
\vspace{1cm}
\psfig{figure=fig2c.epsi,height=7.5cm}
\caption{
Nucleosynthesis in hypernovae and normal core-collapse supernovae for
explosion energies of $E =$ 30 (left) and 1 (right) $ \times$ $
10^{51}$ ergs.  The progenitor model is the 40$M_{\odot}$ star 
model (Nakamura et al. 2001b).
\label{fig:donehn}}
\end{center}
\end{figure}

\begin{figure}
\begin{center}\leavevmode
\psfig{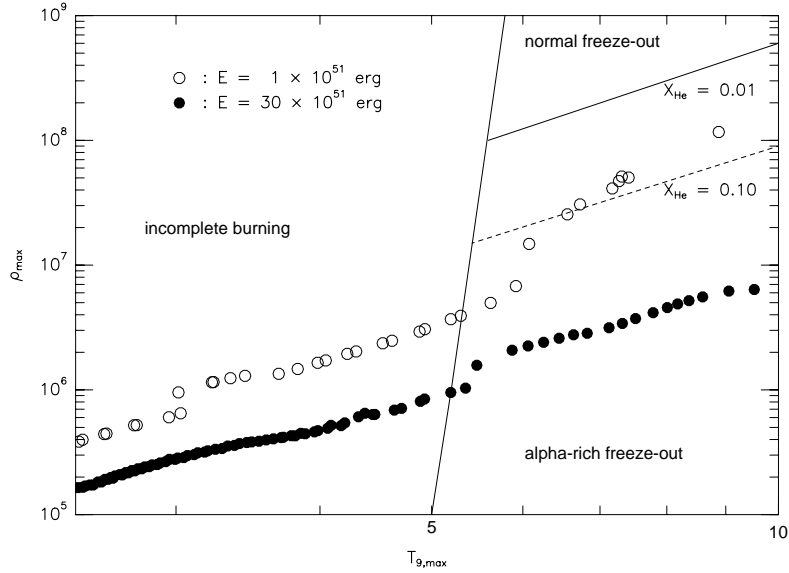}
\caption{
The $\rho$ - $T$ conditions of individual mass zones at their
temperature maximum in hypernovae ($E=$ 30 $ \times$ $ 10^{51}$ ergs:
filled circles) and normal supernovae ($E=$ 1 $ \times$ $ 10^{51}$erg:
open circles) (Nakamura et al. 2001b).
\label{fig:maxrt}}
\end{center}
\end{figure}

In core-collapse supernovae/hypernovae, stellar material undergoes
shock heating and subsequent explosive nucleosynthesis. Iron-peak
elements including Cr, Mn, Co, and Zn are produced in two distinct
regions, which are characterized by the peak temperature, $T_{\rm
peak}$, of the shocked material (Fig. \ref{fig:maxrt}).  For $T_{\rm
peak} > 5\times 10^9$K, material undergoes complete Si burning whose
products include Co, Zn, V, and some Cr after radioactive decays.  For
$4\times 10^9$K $<T_{\rm peak} < 5\times 10^9$K, incomplete Si burning
takes place and its after decay products include Cr and Mn (e.g.,
Hashimoto, Nomoto, Shigeyama 1989; Woosley \& Weaver 1995; Thielemann,
Nomoto \& Hashimoto 1996).

Figure \ref{fig:donehn} shows nucleosynthesis in hypernovae and normal
supernovae for $E_{51} =$ 30 and 1.  The progenitor is 40$M_{\odot}$
star model (Nakamura et al. 2001b).  From this figure, we note the
following characteristics of nucleosynthesis with very large explosion
energies.

1) The region of complete Si-burning, where $^{56}$Ni is dominantly
produced, is extended to the outer, lower density region.  The large
amount of $^{56}$Ni observed in hypernovae (e.g., $ \sim$ 0.4 - 0.7
$M_{\odot}$ for SN1998bw and $ \sim 0.15 M_{\odot}$ for SN1997ef)
implies that the mass cut is rather deep, so that the elements
synthesized in this region such as $^{59}$Cu, $^{63}$Zn, and $^{64}$Ge
(which decay into $^{59}$Co, $^{63}$Cu, and $^{64}$Zn, respectively)
are ejected more abundantly than in normal supernovae.  In the
complete Si-burning region of hypernovae, elements produced by
$\alpha$-rich freezeout are enhanced because nucleosynthesis proceeds
at lower densities than in normal supernovae (Fig. \ref{fig:maxrt}).  
Figure \ref{fig:donehn} clearly shows the trend that a larger amount
of $^{4}$He is left in hypernovae.  Hence, elements synthesized
through capturing of $\alpha$-particles, such as $^{44}$Ti, $^{48}$Cr,
and $^{64}$Ge (decaying into $^{44}$Ca, $^{48}$Ti, and $^{64}$Zn,
respectively) are more abundant.

2) More energetic explosions form a broader incomplete
Si-burning region.  The elements produced mainly in this region such
as $^{52}$Fe, $^{55}$Co, and $^{51}$Mn (decaying into $^{52}$Cr,
$^{55}$Mn, and $^{51}$V, respectively) are synthesized more abundantly
with the larger explosion energy.

3) Oxygen and carbon burning takes place in more extended, lower
density regions for the larger explosion energy.  Therefore, the
elements O, C, Al are burned more efficiently and the abundances of
the elements in the ejecta are smaller, while a larger amount of
burning products such as Si, S, and Ar is synthesized by oxygen
burning.

\begin{figure}
\begin{center}\leavevmode
\psfig{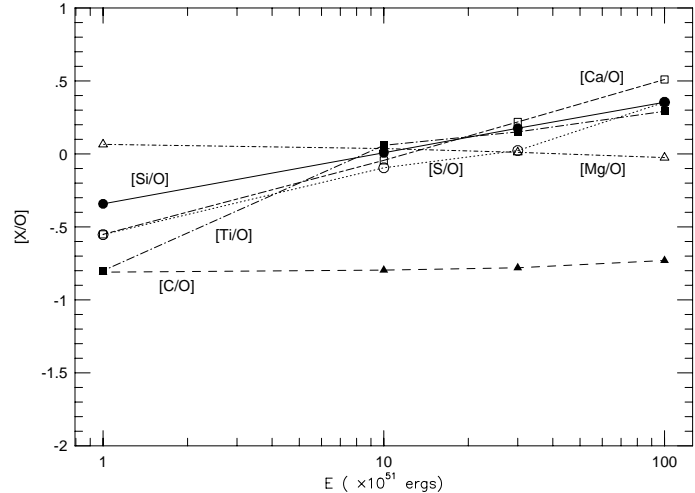}
\caption{
Abundance ratios to oxygen relative to solar values as a function of
the explosion energy $E$.  The progenitor model is the 
25$M_{\odot}$ star model (Nakamura et al. 2001b).
\label{fig:ratiohn}}
\end{center}
\end{figure}

Nucleosynthesis is characterized by large abundance ratios of
intermediate mass nuclei and heavy nuclei with respect to $^{56}$Fe
for more energetic explosions, except for the elements O, C, Al which
are consumed in oxygen and carbon burning.  In particularly, the
amounts of $^{44}$Ca and $^{48}$Ti are increased significantly with
increasing explosion energy because of the lower density regions which
experience complete Si-burning through $\alpha$-rich freezeout.  To
see this more clearly, we add the isotopes and show in Figure
\ref{fig:ratiohn} their ratios to oxygen relative to the solar values
for various explosion energies.  We note that [C/O] and [Mg/O] are not
sensitive to the explosion energy because C, O, and Mg are consumed by
oxygen burning.  On the other hand, [Si/O], [S/O], [Ti/O], and [Ca/O]
are larger for larger explosion energies because Si and S are produced
in the oxygen burning region, and Ti and Ca are increased in the
enhanced $\alpha$-rich freezeout.

\section {Aspherical Explosions}

Despite the success of the hypernova model in reproducing the observed
features of SN 1998bw at early times, some properties of the observed
light curve and spectra at late times are difficult to explain in the
context of a spherically symmetric model.  (1) The decline of the
observed light curve tail is slower than that of the synthetic curve,
indicating that at advanced epochs $\gamma$-ray trapping is more
efficient than expected (Nakamura et al. 2001a; Sollerman et
al. 2000).  (2) In the nebular epoch, the OI]6300\AA\ emission is
narrower than the emission centered at 5200\AA. This latter feature
can be identified as FeII], although it may also contain allowed FeII
transitions. If the identification is correct, the difference in width
may suggest that the mean expansion velocity of iron is higher than
that of oxygen and that a significant amount of oxygen exists at low
velocity (Danziger et al. 1999; Nomoto et al. 2000; Patat et
al. 2001).  Both these features are in conflict with what is expected
from a spherically symmetric explosion model, where $\gamma$-ray
deposition is a decreasing function of time and where iron is produced
in the deepest layers and thus has a lower average velocity than
oxygen. We suggest that these are the signatures of asymmetry in the
ejecta. We thus examine aspherical explosion
models for hypernovae.

On the theoretical side, MacFadyen \& Woosley (1999) showed that the
collapse of a rotating massive core can form a black hole with an
accretion disk, and that a jet emerges along the rotation axis.  The
jet can produce a highly asymmetric supernova explosion (Khokhlov et
al. 1999).  However, these studies did not calculate explosive
nucleosynthesis, nor spectroscopic and photometric features of
aspherical explosions.  Nagataki et al. (1997) performed
nucleosynthesis calculations for aspherical supernova explosions to
explain some features of SN 1987A, but they only addressed the case of
a normal explosion energy.

Maeda et al. (2000) have examined the effect of aspherical (jet-like)
explosions on nucleosynthesis in hypernovae.  They have investigated
the degree of asphericity in the ejecta of SN 1998bw, which is
critically important information to confirm the SN/GRB connection, by
computing synthetic spectra for the various models viewed from
different orientations and comparing the results with the observed
late time spectra (Patat et al. 2001).

\begin{figure}
\begin{center}\leavevmode
\psfig{figure=figure1.epsi,height=6.8cm}
\caption{The isotopic composition of the ejecta in the direction of 
the jet (upper panel) and perpendicular to it (lower panel).
The ordinate indicates the initial spherical Lagrangian coordinate ($M_r$) 
of the test particles (lower scale), and 
the final expansion velocities ($V$) of those particles (upper scale)
(Maeda et al. 2000).
\label{fig:nuc1d}}
\end{center}

\vspace{0.5cm}

\begin{center}\leavevmode
\psfig{figure=figure2_CL.epsi,height=7.4cm}
\caption{The distribution of $^{56}$Ni (open circles) and $^{16}$O (dots).
The open circles and the dots denote test particles in which the mass fraction 
of $^{56}$Ni and $^{16}$O, respectively, exceeds 0.1. 
The lines are density contours at the level of 0.5 (solid), 0.3 (dashed),
0.1 (dash-dotted), and 0.01 (dotted) of the max density, respectively
(Maeda et al. 2000).
\label{fig:nuc2d}}
\end{center}
\end{figure}

\subsection{Nucleosynthesis}

Maeda et al. (2000) have constructed several asymmetric explosion
models.  The progenitor model is the 16 $M_\odot$ He core of the 40
$M_\odot$ star.  The explosion energy is $E$ = 1 $\times$ 10$^{52}$ergs.

Figure~\ref{fig:nuc1d} shows the isotopic composition of the ejecta of
asymmetric explosion model in the direction of the jet (upper panel)
and perpendicular to it (lower panel). In the $z$-direction,
where the ejecta carry more kinetic energy, the shock is stronger and
post-shock temperatures are higher, so that explosive nucleosynthesis
takes place in more extended, lower density regions compared with the
$r$-direction. Therefore, larger amounts of $\alpha$-rich freeze-out
elements, such as $^4$He and $^{56}$Ni (which decays into $^{56}$Fe
via $^{56}$Co) are produced in the $z$-direction than in the
$r$-direction.  Also, the expansion velocity of newly synthesized
heavy elements is much higher in the $z$-direction.

Comparison with the nucleosynthesis of spherically symmetric hypernova
explosions (Nakamura et al. 2001b) shows that the distribution and
expansion velocity of newly synthesized elements ejected in the
$z$-direction is similar to the result of the spherical explosion
model with $E \sim 3 \times 10^{52}$ ergs, although the integrated
kinetic energy is only $E = 1 \times 10^{52}$ ergs.

On the other hand, along the $r$-direction $^{56}$Ni is produced only
in the deepest layers, and the elements ejected in this direction are
mostly the products of hydrostatic nuclear burning stages (O) with
some explosive oxygen-burning products (Si, S, etc). The expansion
velocities are much lower than in the $z$-direction.

Figure~\ref{fig:nuc2d} shows the 2D distribution of $^{56}$Ni and
$^{16}$O in the homologous expansion phase. In the direction closer to
the $z$-axis, the shock is stronger and a low density, $^4$He-rich
region is produced. $^{56}$Ni is distributed preferentially in the jet
direction.  The distribution of heavy elements is elongated in the
$z$-direction, while the $^{16}$O distribution is less aspherical.

\subsection{Nebula Spectra of SN 1998bw}

Mazzali et al. (2001) calculated synthetic nebular-phase spectra of
SN~1998bw using a spherically symmetric NLTE nebular code, based on
the deposition of $\gamma$-rays from $^{56}$Co decay in a nebula of
uniform density and composition. They showed that the nebular lines of
different elements can be reproduced if different velocities are
assumed for these elements, and that a significant amount of
slowly-moving O is necessary to explain the zero velocity peak of the
O {\small I}] line. Different amounts of $^{56}$Ni (0.35$M_\odot$ and
0.6$M_\odot$, respectively) are required to fit the narrow O {\small
I}] line and the broad-line Fe spectrum (see figures in Danziger et
al. 1999 and Nomoto et al. 2000).  This suggests that an aspherical
distribution of Fe and O may explain the observations.

In order to verify the observable consequences of an axisymmetric
explosion, we calculated the profiles of the Fe-dominated blend near
5200\AA, and of O {\small I}] 6300, 6363\AA.  These are the lines
that deviate most from the expectations from a spherically symmetric
explosion.  Line emissivities were obtained from the 1D NLTE code, and
the column densities of the various elements along different lines of
sight were derived from the element distribution obtained from our 2D
explosion models.

\begin{figure}
\begin{center}\leavevmode
\psfig{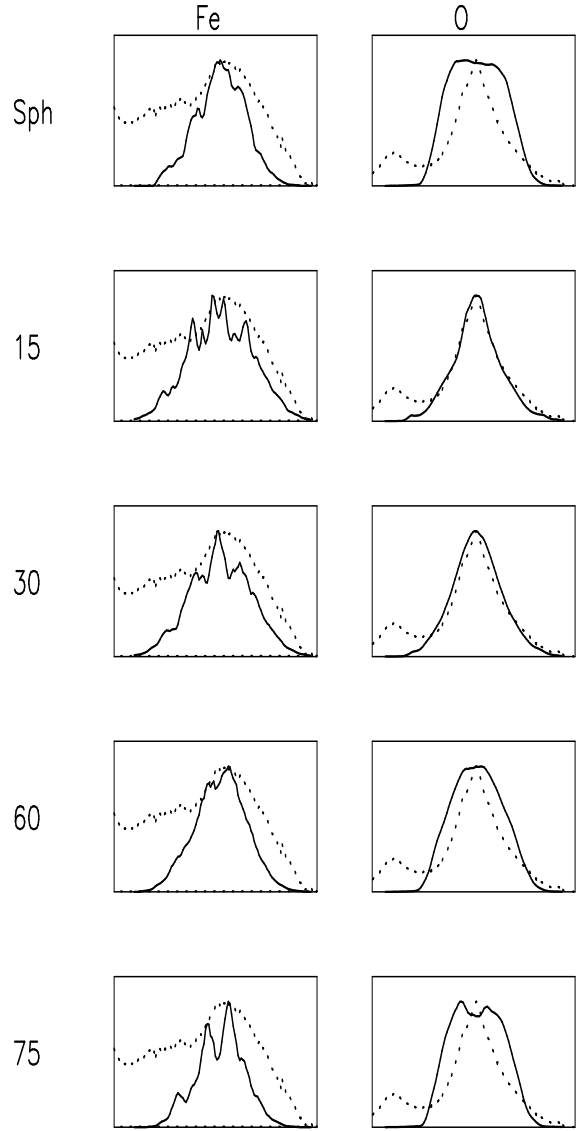}
\caption{The profiles of the Fe-blend (left panels) and of  
O {\small I}] 6300, 6363 \AA\ (right panels) 
viewed at 15$^{\circ}$, 30$^{\circ}$, 60$^{\circ}$ and
75$^{\circ}$ from the jet direction (Maeda 2001).  The top panel shows
the profiles of the spherically symmetric model.
The observed lines at a SN rest-frame epoch of 139 days 
are also plotted for 
comparison (dotted lines, Patat et al. 2000).
The intensities of the strongest lines, normalized to 
O {\small I}] 6300.3\AA\ are: 
FeII\ 5158.8\AA: 0.207; 
FeII\ 5220.1\AA: 0.043; 
FeII\ 5261.6\AA: 0.139; 
FeII\ 5273.3\AA: 0.066; 
FeII\ 5333.6\AA: 0.099; 
FeIII\ 5270.4\AA: 0.086; and 
 O{\small I} 6363.8\AA: 0.331.
\label{fig:line}}
\end{center}
\end{figure}

The nebular line profiles of iron and oxygen with various view angles
are shown in Figure~\ref{fig:line} together with the spherical model
(Maeda 2001).  In this figure, the observed spectrum on day 139 (Patat
et al. 2001) is also plotted for comparison.

The `full-width-at-half-maximum' of our synthetic lines should
be compared to the observational values, which are 440\AA\ for the
Fe-blend and 200\AA\ for the O {\small I}] doublet, and were estimated
from the late time spectrum of 12 Sept 1998, 139 days after the
explosion in the SN rest frame, assuming that the continuum level is
the value around 5700 \AA\ where the flux has the minimum value.

The observed line profiles are not explained with a spherical
model. In a spherical explosion, oxygen is located at higher
velocities than iron, and although the Fe-blend can be wider than the
O line if the O and Fe regions are mixed extensively, the expected
ratio of the width of the Fe-blend and the O line even in fully mixed
model is $\sim 3:2$ (Mazzali et al. 2001). However, the observed ratio
is even larger, $\sim 2:1$.

Because Fe is distributed preferentially along the jet direction, our
aspherical explosion models can produce a larger ratio. If we view the
aspherical explosion model from a near-jet direction, the ratio
between the Fe and O line widths is comparable to the observed value.
For a larger explosion energy, the width of the oxygen line is too
large to be compatible with the observations.

The aspherical explosion models can produce a larger ratio than the
spherical model because Fe is distributed preferentially along the jet
direction.  If the explosion energy is too large, the width of the
oxygen line is too large to be compatible with the observations.

The iron and oxygen profiles viewed at an angle of 15$^{\circ}$ from
the jet direction are compared to the observed spectrum on day 139 in
Figure ~\ref{fig:line}. When the degree of asphericity is high and the
viewing angle is close to the jet direction, the component iron lines
in the blend have double-peaked profiles, the blue- and red-shifted
peaks corresponding to Fe-dominated matter moving towards and away
from us, respectively. This is because Fe is produced mostly along the
$z$ (jet) direction. Because of the high velocity of Fe, the peaks are
widely separated, making the blend wide. This is the case for the
synthetic Fe-blend shown in Figure~\ref{fig:line}.  In contrast, the
oxygen line is narrower and has a sharper peak, because O is produced
mostly in the $r$-direction, at lower velocities and with a less
aspherical distribution.

The Fe-blend line computed with the aspherical model shows some small
peaks (Fig. \ref{fig:line}), which are not seen in the observations. 
However, the detailed profile of the Fe-blend is very sensitive to the
matter distribution.  Mixing of the ejecta may distribute the
fast-moving $^{56}$Ni to lower velocities, which would reduce the
double-peaked profiles of the Fe lines.

\section{Contribution of Hypernovae to the Galactic Chemical Evolution}

\begin{figure}
\begin{center}\leavevmode
\psfig{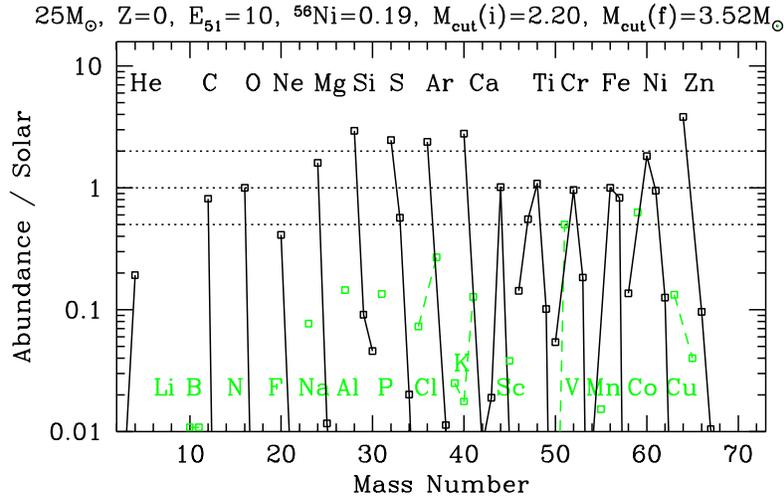}
\caption{Abundance pattern in the ejecta normalized by the solar
$^{16}$O abundances for the 25$M_\odot$ model with $E_{51}=10$.  The
mass-cuts are chosen to give large [Zn/Fe] and [O/Fe]=0 (Umeda \&
Nomoto 2001).
\label{fig:25pop3}}
\end{center}
\end{figure}

The abundance pattern of metal-poor stars with [Fe/H] $< -2$ provides
us with very important information on the formation, evolution, and
explosions of massive stars in the early evolution of the galaxy.
Contributions of hypernova nucleosynthesis to the Galactic chemical
evolution could be important because of their large amounts of heavy
elements production.  Figure \ref{fig:25pop3} shows the abundance
pattern of the explosion of the Pop III (metal-free) 25 $M_\odot$
star.

If hypernovae occurred in the early stage of the Galactic evolution,
the abundance pattern of a hypernova may be observable in some
low-mass halo stars.  This results from the very metal-poor
environment, where the heavy elements synthesized in a single
hypernova (or a single supernova) dominate the heavy element abundance
pattern (Audouze \& Silk 1995).  It is plausible that hypernova
explosions induce star formations.  The low-mass stars produced by
this event should have the hypernova-like abundance pattern and still
exist in the Galactic halo.  The metallicity of such stars is likely
to be determined by the ratio of ejected iron mass from the relevant
hypernova to the mass of hydrogen gathered by the hypernova, which
might be in the range range of $ -4 \lsim$ [Fe/H] $\lsim -2.5$ (Ryan
et al. 1996; Shigeyama \& Tsujimoto 1998; Nakamura et al. 1999; Argast
et al. 2000).

\subsection{Iron, Titanium}

One of the most significant feature of hypernova nucleosynthesis is a
large amount of Fe.  One hypernova can produce 2 - 10 times more Fe
than normal core-collapse supernovae, which is almost the same amount
of Fe as produced in a SN Ia.  This large iron production leads to
small ratios of $\alpha$ elements over iron in hypernovae (Figure
\ref{fig:25pop3}).  In this connection, the abundance pattern of the
very metal-poor binary CS22873-139 ([Fe/H] = $-3.4$) is interesting.
This binary has only an upper limit to [Sr/Fe] $< -1.5$, and therefore
was suggested to be a second generation star (Spite et al. 2000).  The
interesting pattern is that this binary shows almost solar Mg/Fe and
Ca/Fe ratios, as would be the case with hypernovae.  Another feature
of CS22873-139 is enhanced Ti/Fe ([Ti/Fe] $\sim + 0.6$), which could
be explained by an (especially, aspherical) hypernova explosion.

It has been pointed out that Ti is deficient in Galactic chemical
evolution models using supernova yields currently available (e.g.,
Timmes et al. 1995; Thielemann et al. 1996), especially at [Fe/H]
$\lsim -1$ when SNe Ia have not contributed to the chemical evolution.  
However, if the contribution from hypernovae to
Galactic chemical evolution is relatively large (or supernovae are
more energetic than the typical value of 1 $\times$ $10^{51}$ erg),
this problem could be relaxed.  As we have seen in the previous
section, the $\alpha$-rich freezeout is enhanced in hypernovae because
nucleosynthesis proceeds under the circumstance of lower densities and
incomplete Si-burning occurs in more extended regions.  Thus,
$^{40}$Ca, $^{44}$Ca, and $^{48}$Ti are produced and could be ejected
into space more abundantly.

\subsection{Manganese, Chromium, Cobalt}

The observed abundances of metal-poor halo stars show quite
interesting pattern.  There are significant differences between the
abundance patterns in the iron-peak elements below and above [Fe/H]$
\sim -2.5$.  For [Fe/H]$\lsim -2.5$, the mean values of [Cr/Fe] and
[Mn/Fe] decrease toward smaller metallicity, while [Co/Fe] increases
(McWilliam et al. 1995; Ryan et al. 1996).  This trend cannot be
explained with the conventional chemical evolution model that uses
previous nucleosynthesis yields.

\subsubsection{Mass Cut Dependence}

\vspace{0.5cm}

\begin{figure}
\begin{center}\leavevmode
\psfig{figure=15z0fig1.epsi,height=6.5cm}
\caption{
Abundance distribution just after supernova explosion for the Pop III
model with mass 15$M_\odot$.  The lines labeled by Cr, Mn, Co and Zn
actually indicate unstable $^{52}$Fe, $^{55}$Co, $^{59}$Cu, and
$^{64}$Ge, respectively, which eventually decay to the labeled
elements (Umeda \& Nomoto 2001).
\label{fig:15z0}}
\end{center}

\vspace{0.5cm}

\begin{center}\leavevmode
\psfig{figure=fig6.epsi,height=6.5cm}
\caption{
[Mn/Fe] vs. [Co/Fe] and [Cr/Fe] for the 40 $M_\odot$ star (16
$M_\odot$ He star) models with $E = (1.0$ - $30) \times$ $ 10^{51}$
ergs, and the 25 $M_\odot$ star (8 $M_\odot$ He star) models with $E =
(1$ - $10) \times$ $10^{51}$ ergs (Nakamura et al. 2001b).
\label{fig:comn}}
\end{center}
\end{figure}

Nakamura et al. (1999) have shown that this trend of decreasing Cr and
Mn with increasing Co is reproduced by decreasing the mass cut between
the ejecta and the collapsed star for the same explosion model. This
is because Co is mostly produced in complete Si-burning regions, while
Mn and Cr are mainly produced in the outer incomplete Si-burning
region (Fig. \ref{fig:15z0}).  If the mass cut is located at smaller
$M_r$, the mass ratio between the complete and incomplete Si-burning
region is larger.  Therefore, mass cuts at smaller $M_r$ increase the
Co fraction but decrease the Mn and Cr fractions in the ejecta.
Nakamura et al. (1999) have also shown that the observed trend with
respect to [Fe/H] may be explained if the mass cut tends to be smaller
in $M_r$ for the larger mass progenitor.

\subsubsection{Energy Dependence}

Nakamura et al. (2001b) have investigated whether the observed trend
of these iron-peak elements in metal-poor stars can be explained with
the the abundance pattern of hypernovae.  In Figure \ref{fig:comn}, we
plot [Mn/Fe] vs. [Co/Fe] and [Cr/Fe] for the 16 $M_\odot$ He star
models with $E = (1$ - $30)$ $\times$ $ 10^{51}$ ergs, and 8 $M_\odot$
He star models with $E = (1 $ - $10)$ $\times$ $ 10^{51}$ ergs. This
figure clearly shows the correlation between [Mn/Fe] and [Cr/Fe], and
anti-correlation between [Mn/Fe] and [Co/Fe], which are the same
trends as observed in the metal-poor stars.

To understand the dependence on $E$, let us compare the models with 1
$ \times$ $ 10^{51}$ and 10 $ \times$ $ 10^{51}$ ergs which have
almost the same mass cut.  In the model with $E =$ 10 $ \times$ $
10^{51}$ ergs, both complete and incomplete Si-burning regions shift
outward in mass compared with $E =$ 1 $ \times$ $ 10^{51}$ ergs
because of a larger explosion energy.  Thus, the model with larger $E$
has a larger mass ratio between the complete and incomplete Si-burning
regions.  (This relation between $E$ and the mass ratio does not hold
if the explosion energy is too large, because incomplete Si-burning
extends so far out that Mn and Cr increase too much to fit the
metal-poor star data.)

In metal-poor stars, [Mn/Fe] increases with [Fe/H].  Hypernova yields
are consistent with this trend if hypernovae with larger $E$ induce
the formation of stars with smaller [Fe/H]. This supposition is
reasonable because the mass of interstellar hydrogen gathered by a
hypernova is roughly proportional to $E$ (Cioffi et al. 1988;
Shigeyama \& Tsujimoto 1998) and the ratio of the ejected iron mass to
$E$ is smaller for hypernovae than for canonical supernovae.

\subsection{Zinc}

\begin{figure}
\begin{center}\leavevmode
\psfig{figure=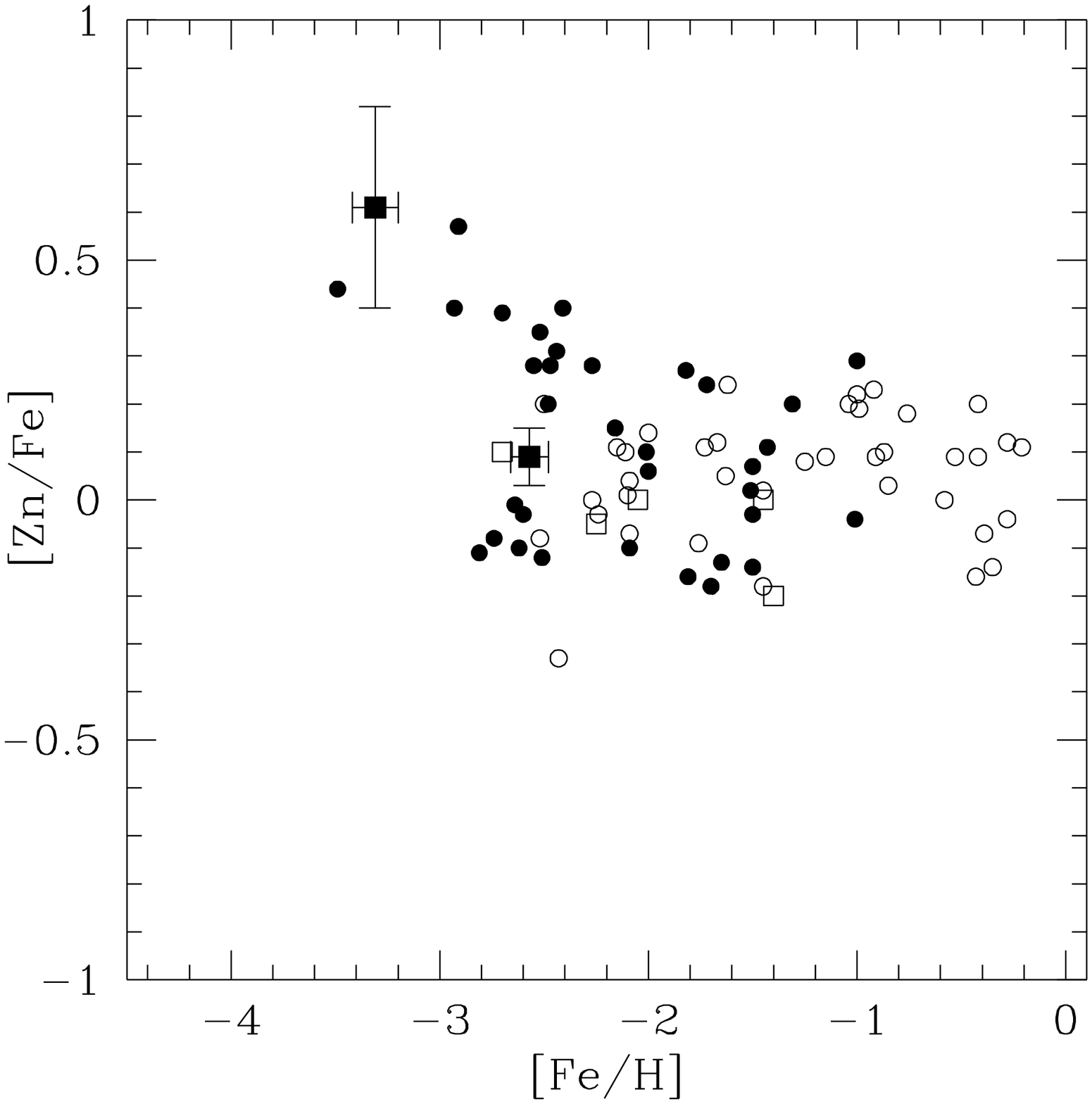,height=7.5cm}
\caption{Observed abundance ratios of [Zn/Fe].  These data are taken
from Primas et al. (2000) (filled circles), Blake et al. (2001)
(filled square) and from Sneden et al. (1991) (others) (Umeda \&
Nomoto 2001).
\label{fig:znfe}}
\end{center}

\vspace{0.5cm}

\begin{center}\leavevmode
\psfig{figure=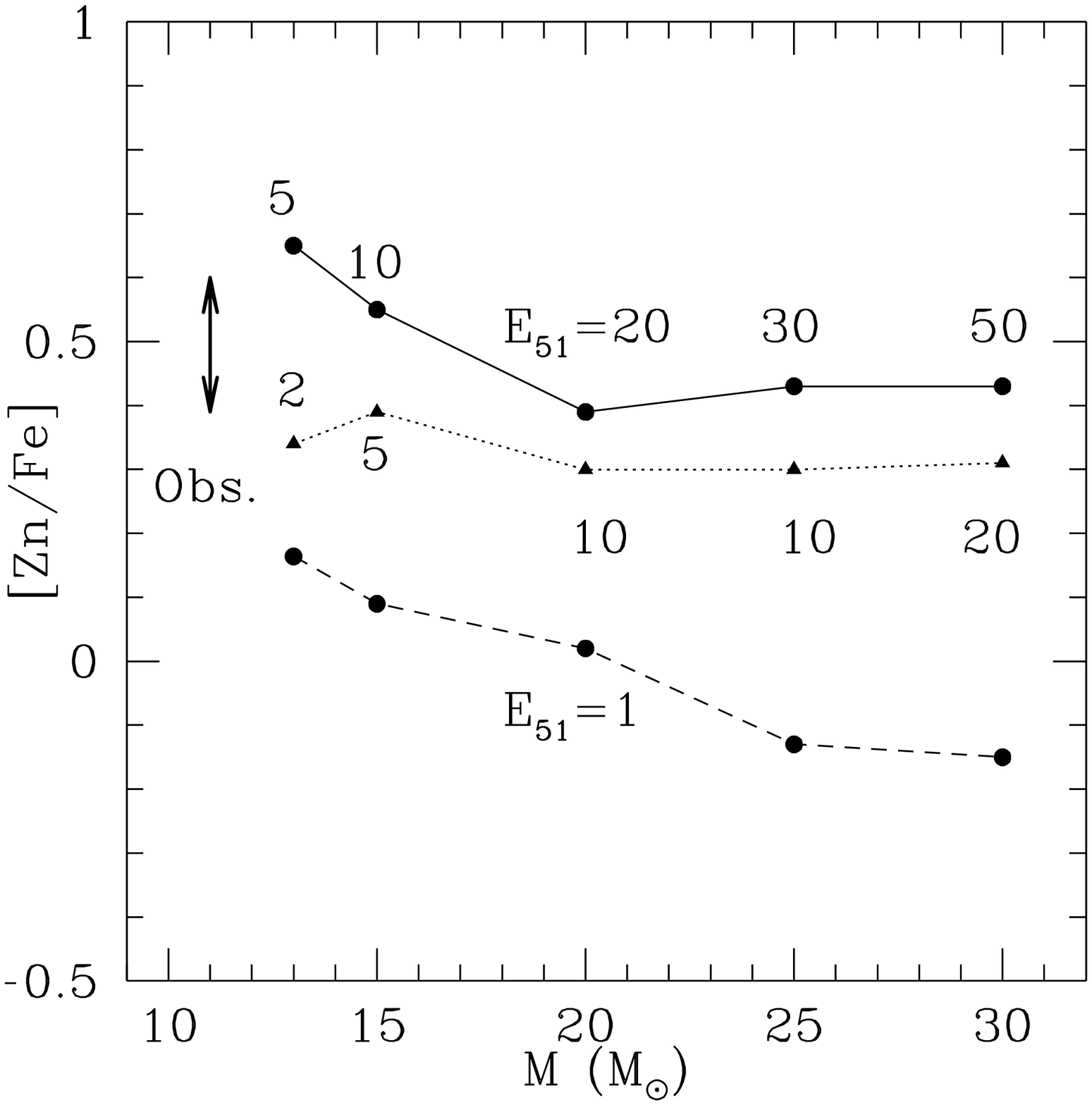,height=7.5cm}
\caption{
The maximum [Zn/Fe] ratios as a function of $M$ and $E_{51}$ (Umeda \&
Nomoto 2001).  The observed large [Zn/Fe] ratio in very low-metal
stars ([Fe/H] $<-2.6$) found in Primas et al. (2000) and Blake et al.
(2001) are represented by a thick arrow.
\label{fig:znme}}
\end{center}
\end{figure}

     For Zn, early observations have shown that [Zn/Fe]$ \sim 0$ for
[Fe/H] $\simeq -3$ to $0$ (Sneden, Gratton, \& Crocker 1991).
Recently Primas et al. (2000) have suggested that [Zn/Fe] increases
toward smaller metallicity as seen in Figure \ref{fig:znfe}, and Blake
et al. (2001) has one with [Zn/Fe] $\simeq 0.6$ at [Fe/H] = $-3.3$
(see Ryan 2001).

     As for Zn, its main production site has not been clearly
identified.  If it is mainly produced by s-processes, the abundance
ratio [Zn/Fe] should decrease with [Fe/H].  This is not consistent
with the observations of [Zn/Fe]$ \sim 0$ for [Fe/H] $\simeq -2.5$ to
$0$ and the increase in [Zn/Fe] toward lower metallicity for
[Fe/H]$\lsim -2.5$.  Another possible site of Zn production is SNe II.
However, previous nucleosynthesis calculations in SNe II appears to
predict decreasing Zn/Fe ratio with Fe/H (Woosley \& Weaver 1995;
Goswami \& Prantzos 2000).

     Understanding the origin of the variation in [Zn/Fe] is very
important especially for studying the abundance of Damped Ly-$\alpha$
systems (DLAs), because [Zn/Fe] $=0$ is usually assumed to determine
their abundance pattern.  In DLAs super-solar [Zn/Fe] ratios have
often been observed, but they have been explained that assuming dust
depletion for Fe is larger than it is for Zn.  However, recent
observations suggest that the assumption [Zn/Fe] $=0$ may not be
always correct.

\subsubsection{Spherical Hypernova Models}

Umeda \& Nomoto (2001) have calculated nucleosynthesis in massive Pop
III stars and explored the parameter ranges ($M_{\rm cut}(i)$, $Y_e$,
$M$, and $E$) to reproduce [Zn/F] $\sim$ 0.5 observed in extremely
metal-poor stars.  Their main results are summarized as follows.

1) The interesting trends of the observed ratios [(Zn, Co, Mn, Cr)/Fe]
can be understood in terms of the variation of the mass ratio between
the complete Si burning region and the incomplete Si burning region.
The large Zn and Co abundances observed in very metal-poor stars are
obtained if the mass-cut is deep enough, or equivalently if deep
material from complete Si-burning region is ejected by mixing or
aspherical effects.  Vanadium also appears to be abundant at low
[Fe/H].  Since V is also produced mainly in the complete Si-burning
region, this trend can be explained in the same way as those of Zn and
Co.

2) The mass of the incomplete Si burning region is sensitive to the
progenitor mass $M$, being smaller for smaller $M$.  Thus [Zn/Fe]
tends to be larger for less massive stars for the same $E$.

3) The production of Zn and Co is sensitive to the value of $Y_e$,
being larger as $Y_e$ is closer to 0.5, especially for the case of
a normal explosion energy ($E_{51}\sim 1$).

4) A large explosion energy $E$ results in the enhancement of the
local mass fractions of Zn and Co, while Cr and Mn are not enhanced.
This is because larger $E$ produces larger entropy and thus
more $\alpha$-rich environment for $\alpha$-rich freeze-out.

5) To be consistent with the observed [O/Fe] $\sim$ 0 - 0.5 as well as
with [Zn/Fe] $\sim$ 0.5 in metal-poor stars, they propose that the
``mixing and fall-back" process or aspherical effects are significant
in the explosion of relatively massive stars.

The dependence of [Zn/Fe] on $M$ and $E$ is summarized in Figure
\ref{fig:znme}.  They have found that models with $E_{51} =$ 1 do not
produce sufficiently large [Zn/Fe].  To be compatible with the
observations of [Zn/Fe] $\sim 0.5$, the explosion energy must be much
larger, i.e., $E_{51} \gsim 2$ for $M \sim 13 M_\odot$ and $E_{51}
\gsim 20$ for $M \gsim 20 M_\odot$.

\begin{figure}
\begin{center}\leavevmode
\psfig{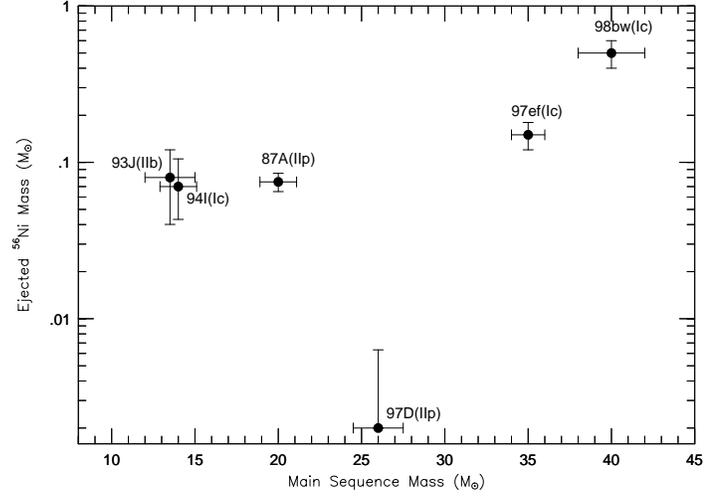}
\caption{
Ejected $^{56}$Ni mass versus the main sequence mass of the
progenitors of several bright supernovae obtained from light curve
models (Iwamoto et al. 2000).
\label{fig:nimass}}
\end{center}
\end{figure}

     Observationally, the requirement of the large $E$ might suggest
that large $M$ stars are responsible for large [Zn/Fe], because $E$
and $M$ can be constrained from the observed brightness and light
curve shape of supernovae as follows (Fig. \ref{fig:nimass}).  The
recent supernovae 1987A, 1993J, and 1994I indicate that the
progenitors of these normal SNe are 13 - 20 $M_\odot$ stars and
$E_{51} \sim$ 1 - 1.5 (Nomoto et al. 1993, 1994; Shigeyama et
al. 1994; Blinnikov et al. 2000).  On the other hand, the masses of
the progenitors of hypernovae with $E_{51} >$ 10 (SNe 1998bw, 1997ef,
and 1997cy) are estimated to be $M \gsim 25 M_\odot$ (Nomoto et
al. 2000; Iwamoto et al. 1998, 2000; Woosley et al. 1999; Turatto et
al. 2000).  This could be related to the stellar mass dependence of
the explosion mechanisms and the formation of compact remnant, i.e.,
less massive stars form neutron stars, while more massive stars tend
to form black holes.

To explain the observed relation between [Zn/Fe] and [Fe/H], we
further need to know how $M$ and $E$ of supernovae and [Fe/H] of
metal-poor halo stars are related.  In the early galactic epoch when
the galaxy is not yet chemically well-mixed, [Fe/H] may well be
determined by mostly a single SN event (Audouze \& Silk 1995).  The
formation of metal-poor stars is supposed to be driven by a supernova
shock, so that [Fe/H] is determined by the ejected Fe mass and the
amount of circumstellar hydrogen swept-up by the shock wave (Ryan et
al. 1996).

Explosions with the following two combinations of $M$ and $E$ can be
responsible for the formation of stars with very small [Fe/H]:

i) Energetic explosions of massive stars ($M \gsim 25 M_\odot$): For
these massive progenitors, the supernova shock wave tends to propagate
further out because of the large explosion energy and large
Str\"omgren sphere of the progenitors (Nakamura et al. 1999).  The
effect of $E$ may be important since the hydrogen mass swept up by the
supernova shock is roughly proportional to $E$ (e.g., Ryan et al 1996;
Shigeyama \& Tsujimoto 1998).

ii) Normal supernova explosions of less massive stars ($M \sim 13
M_\odot$): These supernovae are assumed to eject a rather small mass
of Fe (Shige-yama \& Tsujimoto 1998), and most SNe are assumed to
explode with normal $E$ irrespective of $M$.

The above relations lead to the following two possible scenarios to
explain [Zn/Fe] $\sim 0.5$ observed in metal-poor stars.

i) Hypernova-like explosions of massive stars ($M \gsim 25
M_\odot$) with $E_{51} > 10$:  Contribution of highly asymmetric
explosions in these stars may also be important.  The question is 
what fraction of such massive stars explode as hypernovae; the
IMF-integrated yields must be consistent with [Zn/Fe] $\sim$ 0 at
[Fe/H] $\gsim -2.5$.

ii) Explosion of less massive stars ($M \lsim 13 M_\odot$) with
$E_{51} \gsim 2$ or a large asymmetry: This scenario, after
integration over the IMF, can reproduce the observed abundance pattern
for [Fe/H]$\gsim -2$.  However, the Fe yield has to be very small in
order to satisfy the observed [O/Fe] value ($\gsim 0.5$) for the
metal-poor stars.  For example, the $^{56}$Ni mass yield of our
13$M_\odot$ model has to be less than 0.006$M_\odot$, which appears to
be inconsistent with the nearby SNe II observations.

It seems that the [O/Fe] ratio of metal-poor stars and the $E$-$M$
relations from supernova observations (Fig. \ref{fig:nimass}) favor
the massive energetic explosion scenario for enhanced [Zn/Fe].
However, we need to construct detailed galactic chemical evolution
models to distinguish between the two scenarios for [Zn/Fe].

\begin{figure}
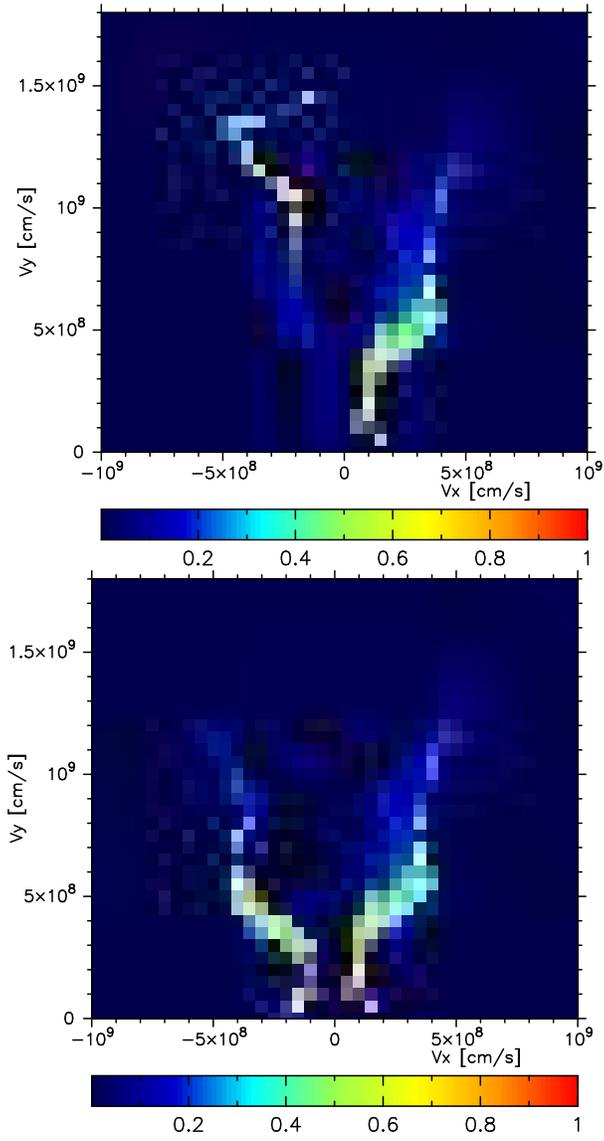

\begin{center}\leavevmode
\psfig{figure=2dzn_CL2.ps,height=7.5cm,angle=270}
\vspace{0.5cm}
\psfig{figure=2dmn_CL2.ps,height=7.5cm,angle=270}
\caption{The density distributions of Zn (top: left half) and
Fe (top: right half) and that of Mn (bottom: left half)
and Fe (bottom: right half). The densities of each element are
represented with linear scale, from 0 to the max density of
each element (Maeda 2001).
\label{fig:2delm}}
\end{center}
\end{figure}

\begin{figure}
\begin{center}\leavevmode
\psfig{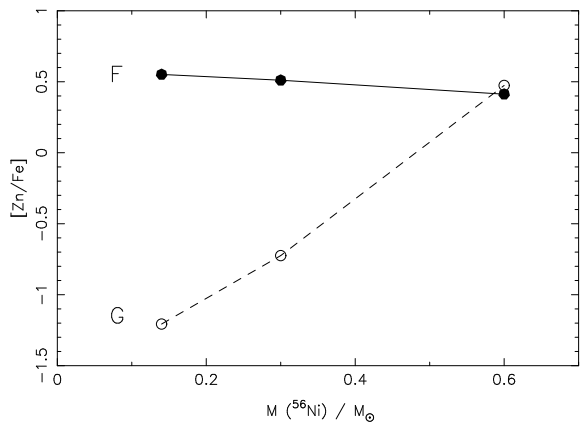}
\caption{[Zn/Fe] as a function of the mass of ejected $^{56}$Ni (Maeda
2001).  Larger $^{56}$Ni mass corresponds to deeper mass cut.  The
filled and open circles denote the aspherical and spherical models,
respectively.  The progenitor is the 40$M_\odot$ star (Umeda \& Nomoto
2001).
\label{fig:znu40}}
\end{center}
\end{figure}

\subsubsection{Aspherical Hypernova Models}

Iron-peak elements are produced in the deep region near the mass cut,
so that their production is strongly affected by asphericity of an
explosion. Figure~\ref{fig:2delm} shows the 2D density distribution of
$^{64}$Ge (which decays into $^{64}$Zn) and that of $^{55}$Co (which
decays into $^{55}$Mn). $^{64}$Zn, which is produced by the strong
$\alpha$-rich freezeout, is distributed preferentially in the
$z$-direction.  Moreover, $^{64}$Zn production is strongly enhanced
compared with a spherical model, since the post-shock temperature
along the $z$-direction is much higher than that of a spherical model. 
On the other hand, $^{55}$Mn, which is produced by incomplete silicon
burning, surrounds $^{56}$Fe and located preferentially in the
$r$-direction. Accordingly, materials with higher [Zn/Fe] and lower
[Mn/Fe] are ejected along the $Z$-direction with higher velocities,
and materials with lower [Zn/Fe] and higher [Mn/Fe] are ejected toward
the $r$-direction with lower velocities.  

Figure \ref{fig:znu40} shows that [Zn/Fe] is larger for deeper mass
cut (i.e., larger $^{56}$Ni mass) for spherical explosion, while
[Zn/Fe] $\sim$ 0.5 is realized irrespective of the mass cut.  This
comes from the difference in the site of the Zn production. In the
spherical case, Zn is produced only in the deepest layer, while in the
aspherical model, the complete silicon burning region is elongated to the
$z$ (jet) direction (Figure {\ref{fig:2delm}), so that [Zn/Fe] is
enhanced irrespective of the mass cut.

In this way, larger asphericity in the explosion leads to larger
[Zn/Fe] and [Co/Fe], but to smaller [Mn/Fe] and [Cr/Fe].  Then, if the
degree of the asphericity tends to be larger for lower [Fe/H], the
trends of [Zn, Co, Mn, Cr/Fe] follow the observed ones.

It is likely that in the real situations, the asphericity, mixing, and
the effects of the mass-cut (Nakamura et al. 1999) work to make
observed trends.  Then, the restrictions encountered by each of them
(e.g., requirement of large $Y_e$, large amounts of mixing) may be
relaxed.

\section{Other Observational Signatures of Hypernova Nucleosynthesis}

\subsection{Starburst Galaxies}

X-ray emissions from the starburst galaxy M82 were observed with ASCA
and the abundances of several heavy elements were obtained (Tsuru et
al. 1997).  Tsuru et al. (1997) found that the overall metallicity of
M82 is quite low, i.e., O/H and Fe/H are only 0.06 - 0.05 times solar,
while Si/H and S/H are $\sim$ 0.40 - 0.47 times solar.  This implies
that the abundance ratios are peculiar, i.e., the ratio O/Fe is about
solar, while the ratios of Si and S relative to O and Fe are as high
as $\sim$ 6 - 8.  These ratios are very different from those ratios in
SNe II.  The age of M82 is estimated to be $\lsim 10^8$ years, which
is too young for Type Ia supernovae to contribute to enhance Fe
relative to O.  Tsuru et al. (1997) also estimated that the explosion
energy required to produce the observed amount of hot plasma per
oxygen mass is significantly larger than that of normal SNe II (here
the oxygen mass dominates the total mass of the heavy elements).
Tsuru et al. (1997) thus concluded that neither SN Ia nor SN II can
reproduce the observed abundance pattern of M82.

\begin{figure}[t]
\begin{center}\leavevmode
\psfig{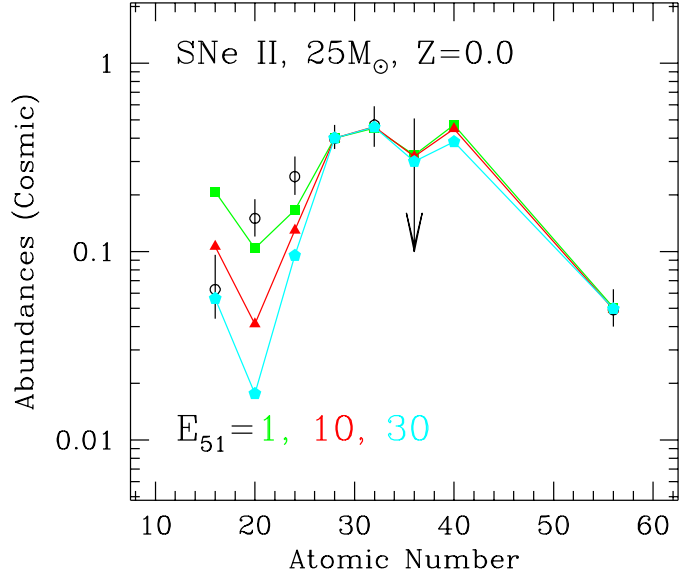}
\caption{
Abundance patterns in the ejecta of 25$M_\odot$ metal-free SN II and
hypernova models compared with abundances (relative to the solar
values) of M82 observed with ASCA (Tsuru et al. 1997).  Here, the open
circles with error bars show the M82 data.  The filled square,
triangle, and pentagons represent $E_{51}$=1, 10, and 30 models,
respectively, where $E_{51}$ is the explosion energy in $10^{51}$
ergs.  Theoretical abundances are normalized to the observed Si data,
and the mass cuts are chosen to eject 0.07, 0.095, and 0.12
($M_\odot$) Fe for $E_{51}$ = 1, 10, and 30, respectively (Umeda et
al. 2001).
\label{fig:m82}}
\end{center}
\end{figure}

Compared with normal SNe II, the important characteristic of hypernova
nucleosynthesis is the large Si/O, S/O, and Fe/O ratios.  Figure
\ref{fig:m82} shows the good agreement between the hypernova model and
the observed abundances in M82 (Umeda et al. 2001).  Hypernovae could
also produce larger $E$ per oxygen mass than normal SNe II.  We
therefore suggest that hypernova explosions may make important
contributions to the metal enrichment and energy input to the
interstellar matter in M82.  If the IMF of the star burst is
relatively flat compared with Salpeter IMF, the contribution of very
massive stars and thus hypernovae could be much larger than in our
Galaxy.

\subsection{Black Hole Binaries}

X-ray Nova Sco (GRO J1655-40), which consists of a massive black hole
and a low mass companion (e.g., Brandt et al. 1995; Nelemans et
al. 2000), also exhibits what could be the nucleosynthesis signature
of a hypernova explosion.  The companion star is enriched with Ti, S,
Si, Mg, and O but not much Fe (Israelian et al. 1999).  This is
compatible with heavy element ejection from a black hole progenitor.
In order to eject large amount of Ti, S, and Si and to have at least
$\sim$ 4 $M_\odot$ below mass cut and thus form a massive black hole,
the explosion would need to be highly energetic (Figure
\ref{fig:donehn}; Israelian et al. 1999; Brown et al. 2000;
Podsiadlowski et al. 2001).  A hypernova explosion with the mass cut
at large $M_r$ ejects a relatively small mass Fe and would be
consistent with these observed abundance features.  

Alternatively, if the explosion which resulted from the formation of
the black hole in Nova Sco was asymmetric, then it is likely that the
companion star captured material ejected in the direction away from
the strong shock which contained relatively little Fe compared with
the ejecta in the strong shock (Maeda et al. 2000).

\section{Summary and Discussion}\label{sec:summary}

We investigated explosive nucleosynthesis in hypernovae.  Detailed
nucleosynthesis calculations were performed for both spherical and
aspherical explosions.  We also studied implications to Galactic
chemical evolution and the abundances in metal-poor stars.  The
characteristics of hypernova yields compared to those of ordinary
core-collapse supernovae are summarized as follows:

1) Complete Si-burning takes place in more extended region, so that
the mass ratio between the complete and incomplete Si burning regions
is generally larger in hypernovae than normal SNe II.  As a result,
higher energy explosions tend to produce larger [(Zn, Co)/Fe], small
[(Mn, Cr)/Fe], and larger [Fe/O].  If hypernovae made significant
contributions to the early Galactic chemical evolution, it could
explain the large Zn and Co abundances and the small Mn and Cr
abundances observed in very metal-poor stars.  The large [Fe/O]
observed in some metal-poor stars and galaxies might possibly be the
indication of hypernovae rather than Type Ia supernovae.

2) In hypernovae, Si-burning takes place in lower density regions, so
that the effects of $\alpha$-rich freezeout is enhanced.  Thus
$^{44}$Ca, $^{48}$Ti, and $^{64}$Zn are produced more abundantly than
in normal supernovae.  The large [(Ti, Zn)/Fe] ratios observed in very metal
poor stars strongly suggest a significant contribution of hypernovae.

3) Oxygen burning also takes place in more extended regions for the
larger explosion energy.  Then a larger amount of Si, S, Ar, and Ca
("Si") are synthesized, which makes the Si/O ratio larger.  The
abundance pattern of the starburst galaxy M82, characterized by
abundant Si and S relative to O and Fe, may be attributed to hypernova
explosions if the IMF is relatively flat, and thus the contribution of
massive stars to the galactic chemical evolution is large.

More direct signature of hypernovae can be seen in the black hole
binary X-ray Nova Sco (GRO J1655-40), which indicates the production
of a large amount of Ti, S, and Si when a massive black hole is formed
as a compact remnant.

We also examine aspherical explosion models.  The explosive shock is
stronger in the jet-direction.  The high explosion energies
concentrated along the jet result in hypernova-like nucleosynthesis.
On the other hand, the ejecta orthogonal to the jet is more like
ordinary supernovae.  Therefore, asphericity in the explosions
strengthens the nucleosynthesis properties of hypernovae as summarized
above in 1) and 2) but not 3).

It is shown that such a large Zn abundance as [Zn/Fe] $\sim$ 0.5
observed in metal-poor stars can be realized in certain supernova
models (Umeda \& Nomoto 2001).  This implies that the assumption of
[Zn/Fe] $\sim 0$ usually adopted in the DLA abundance analyses may not
be well justified.  Rather [Zn/Fe] may provide important information
on the IMF and/or the age of the DLA systems.

Properties of hypernova nucleosynthesis suggest that hypernovae of
massive stars may make important contributions to the Galactic (and
cosmic) chemical evolution, especially in the early low metallicity
phase.  This may be consistent with the suggestion that the IMF of Pop
III stars is different from that of Pop I and II stars, and that more
massive stars are abundant for Pop III (e.g., Nakamura \& Umemura
1999; Omukai \& Nishi 1999; Bromm, Coppi \& Larson 1999).

\begin{acknowledgments}
We would like to thank P. Mazzali, N. Patat, S. Ryan, C. Kobayashi, M. 
Shirouzu, F. Thielemann for useful discussion.  This work has been
supported in part by the grant-in-Aid for Scientific Research
(07CE2002, 12640233) of the Ministry of Education, Science, Culture,
and Sports in Japan.

\end{acknowledgments}
%------------ end of article ------------------->>

%% optional
%\section{Summary}

%% appendix optional
%\appendix{This is the Appendix Title}
%This is an appendix with a title.

%\appendix{}
%This is an appendix without a title.

%
% Bibliography made with BibTeX:
%% apalike is preferred if you have used \kluwerbib, above.
%% Otherwise you may use any .bst style your editor approves.

%This will allow many Bib\TeX\ bibliographies in one book.
%See the documentation, edbk.doc, for more information.

%\bibliographystyle{apalike}
%\chapbblname{<name of .bbl file>}
%\chapbibliography{<name of .bib file>}

%or 

\end{document}